\documentclass{iau} 
\usepackage{graphicx}

\title[IAUS~337---Ephemerides and Timing]%
{Solar System Ephemerides, Pulsar Timing, Gravitational Waves, \&
	Navigation}

\author[Lazio et al.]%
{T.~Joseph~W.~Lazio${}^1$,
S.~Bhaskaran${}^1$, C.~Cutler${}^1$, W.~M.~Folkner${}^1$, R.~S.~Park${}^1$,
J.~A.~Ellis${}^2$, T.~Ely${}^1$, S.~R.~Taylor${}^3$, M.~Vallisneri${}^1$}
\affiliation{$^1$Jet Propulsion Laboratory, California Institute of
Technology, 4800 Oak Grove Dr, Pasadena, CA  91109 \hbox{USA}; %
$^2$Department of Physics \& Astronomy, West Virginia University,
	Morgantown, WV 26506 \hbox{USA}; %
${}^3$California Institute of Technology, Pasadena, CA  91125  USA%
}

\pubyear{2017}
\volume{337}  
\jname{Pulsar Astrophysics -­ The Next 50 Years}
\editors{P.~Weltevrede, B.~B.~P.~Perera, L.~Levin Preston \& S.~Sanidas, eds.}

\begin{document}

\maketitle

\begin{abstract}
In-spiraling supermassive black holes should emit gravitational waves,
which would produce characteristic distortions in the time of arrival
residuals from millisecond pulsars.  Multiple national and regional
consortia have constructed pulsar timing arrays by precise timing of
different sets of millisecond pulsars.  An essential aspect of
precision timing is the transfer of the times of arrival to a
(quasi-)inertial frame, conventionally the solar system
barycenter.  The barycenter is determined from the
knowledge of the planetary masses and orbits, which has been refined
over the past 50~years by multiple spacecraft.  Within the North
American Nanohertz Observatory for Gravitational Waves (NANOGrav),
uncertainties on the solar system barycenter are emerging as an
important element of the NANOGrav noise budget.  We describe
what is known about the solar system barycenter, touch upon how
uncertainties in it affect gravitational wave studies with pulsar
timing arrays, and consider future trends in spacecraft navigation.
\end{abstract}

\firstsection

\section{The Solar System Ephemeris}\label{sec:lazio.ssb}

The timing of ``Pulsar Astrophysics --­ The Next 50~Years'' coincided
not only with the $50^{\mathrm{th}}$ anniversary of Dame Jocelyn
Bell-Burnell's efforts to understand ``scruff,'' it coincided with the
$40^{\mathrm{th}}$ anniversary for the Voyager~1 and~2 launches
(1977 September~5 and August~20, respectively).  The Voyager
spacecraft revolutionized our understanding of the solar system with
their flybys of Jupiter, Saturn, Uranus (Voyager~2), and Neptune
(Voyager~2).

\begin{figure}
\centering
\includegraphics[width=0.75\textwidth]{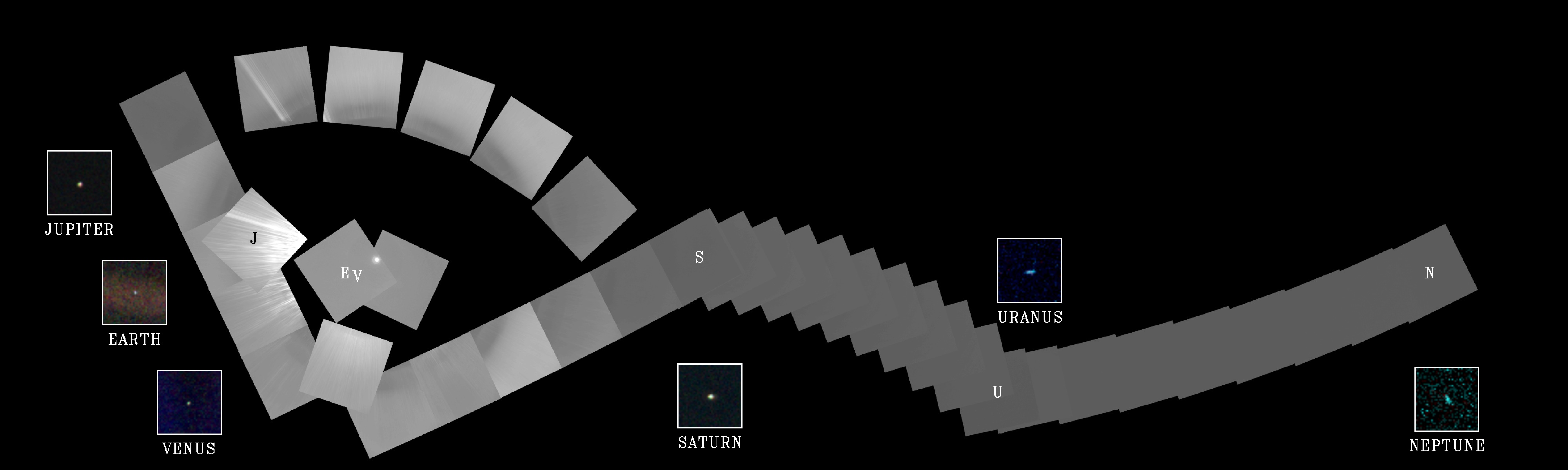}\hspace*{-25ex}%
\includegraphics[width=0.5\textwidth]{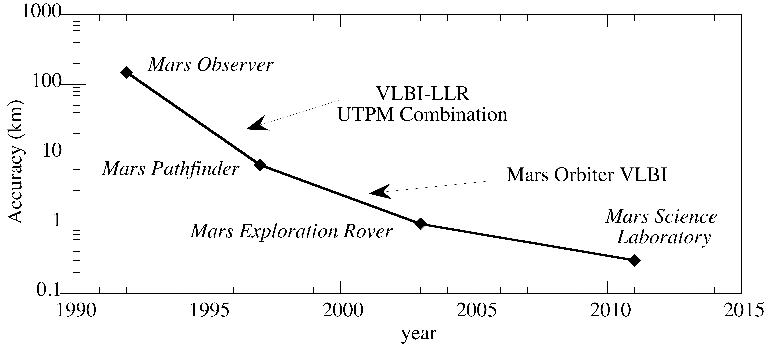}
\caption{(\textit{Left}) Voyager~1 Family Portrait showing most of the planets as seen from the edge of the solar system.  Navigating
the Voyager spacecraft, and subsequent spacecraft, on their
trajectories and knowing the orbits of the planets sufficient to
obtain the Family Portrait has required improved knowledge of the
solar system ephemeris over the past 50~years.  (Credit: \hbox{NASA}/JPL-Caltech)
(\textit{Right}) Improvement in the accuracy of navigation to Mars
over the past nearly 30~years.  While specific to Mars, a similar
trend holds for navigation throughout the solar system.}
\label{fig:lazio.portrait}
\end{figure}

Among the iconic images from the Voyager spacecraft is the ``Family
Portrait'' (Figure~\ref{fig:lazio.portrait}).  The last images
acquired by Voyager~1 before its camera was turned off to save power,
the Family Portrait shows most of the planets as seen from the edge of
the solar system.  Obtaining it required accurate knowledge of the
solar system ephemeris---the masses and orbits of the planets and
minor bodies---both to navigate the Voyager spacecraft on their
journeys and to know where to point the Voyager~1 camera.

Over the past 50~years, at least one spacecraft has flown past each of
the planets (with multiple minor bodies), and
at least one spacecraft has orbited most of the planets.  These
missions have been enabled by continual improvements in our knowledge
of the solar system ephemeris and navigation techniques
(Figure~\ref{fig:lazio.portrait}).  Today, the orbits of the inner
planets are known to a few meters, aided by the multiple orbiters at
both Venus and Mars and their relatively short orbital periods.  In
the outer solar system, orbits are less well known, due to the fewer
number of spacecraft that have visited those planets and the (much)
longer orbital periods; the Saturnian orbit is the most well
determined (tens of meters) due to the recently concluded
\textit{Cassini} mission.

\section{Gravitational Waves, Pulsar Timing, and Solar System Ephemerides}\label{sec:lazio.timing}

Precision timing involves the transformation of the pulse time
of arrival at a telescope located on the Earth into a (quasi-)inertial
frame, typically taken to be the solar system barycenter (Lorimer \&
Kramer~2004).  Among the corrections is one for the Roemer delay,
\begin{equation}
\Delta t_R 
 = \frac{\mathbf{r}_{\mathrm{SSB}} \cdot \mathbf{\hat n}}{c},
\label{eqn:lazio.roemer}
\end{equation}
where $\mathbf{r}_{\mathrm{SSB}}$ is the vector between the Earth and
the solar system barycenter, $\mathbf{\hat n}$ is the unit vector in
the direction of pulsar, and~$c$ is the speed of
light.  With ${r}_{\mathrm{SSB}} \sim 1$~au, $\Delta t_R \approx 500$~s.

Detecting low frequency ($f \sim 10$~nHz) gravitational waves (GWs)
has emerged as an increasing focus for precision pulsar timing among
national and regional consortia.  It is reasonably well established
that most major galaxies host central supermassive black holes
(SMBHs), and the merger of galaxies should result in the two SMBHs
falling to the center of the merger product, under the influence of
dynamical friction (e.g., Begelman et al.~1980; Khan et al.~2016).
The two SMBHs should form a binary, which in most scenarios, begins to
radiate GWs and harden, with the two SMBHs eventually merging (e.g.,
Sesana~2013).  The ensemble of SMBH binaries should produce a GW
background.  Initial expectations were that the GW background would be
isotropic (e.g., Jaffe \& Backer~2003), but recent work has addressed
whether individual binaries could produce an anisotropic GW background
or even be detectable (e.g., Mingarelli et al.~2017).

The expected magnitude for pulse arrival time distortions due to low
frequency GWs is $\Delta t_{\mathrm{GW}} \sim 10$~ns.  Clearly,
detecting low frequency GWs requires knowledge (and control, where
possible) of the various contributions to the timing ``noise budget.''
The connection between the knowledge of the solar system ephemeris and
GW detection is now clear.  Ideally, uncertainty in the barycenter
should be $\sigma_{\mathrm{SSB}} < \Delta t_{\mathrm{GW}}
\sim 10\,\mathrm{ns}$, corresponding to knowledge of the position of
the barycenter to of order a few meters.

Unfortunately, knowledge of the barycenter at this precision does not
exist.  The dominant contribution to the uncertainty results from the
outer solar system, notably from Jupiter, Uranus, and Neptune.  The
\textit{Cassini} mission improved the knowledge of Saturn's mass and
orbit, and the Juno mission is expected to provide similar
improvements for Jupiter, but no orbiter has visited Uranus or
Neptune.  Moreover, uncertainties in their masses and orbits are
degenerate with uncertainties in the orbit of Jupiter, which is the
dominant contribution to estimating the position of the barycenter.

\begin{figure}
\centering
\includegraphics[width=0.75\textwidth]{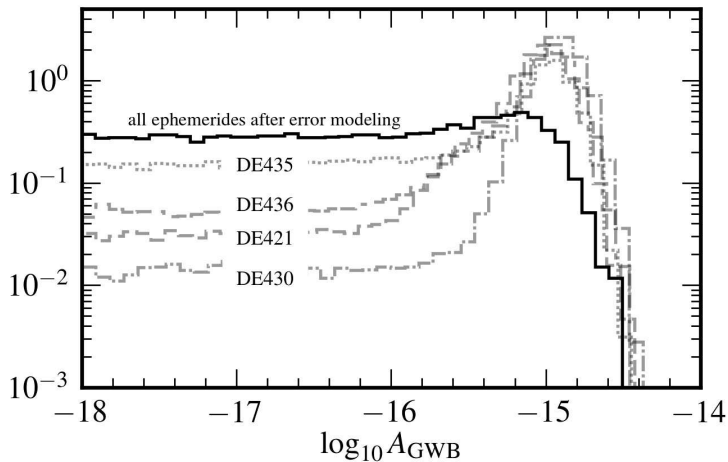}
\vspace*{-4ex}
\caption{Illustration of the effect of solar system ephemerides
uncertainties on gravitational wave detection, from the forthcoming
NANOGrav 11~Year analysis.  Different ephemerides produce different
posterior distributions (as labeled) of the stochastic GW background
amplitude, for a power law in characteristic strain ($h[f] \propto
A_{\mathrm{GWB}}f^\alpha$).  The solid curve shows the resulting
posteriors, which are essentially indistinguishable, if the masses of
the outer planets and the orbital parameters of Jupiter are included
in the GW analysis.  The value of the posterior at small
$\log_{10}A_{\mathrm{GWB}}$ is proportional to the Bayes ratio for the
data favoring a model with no GW background to a model with a GW
background.  The differences between the curves are apparent and
demonstrate that not accounting properly for uncertainties in the
ephemerides could result in an erroneous detection or missing a true
detection.}
\label{fig:lazio.gw}
\end{figure}

Within the North American Nanohertz Observatory for Gravitational
Waves (NANOGrav), this uncertainty in the knowledge of the ephemeris
is being taken into account in the GW analysis.  Modeling of time of
arrival residuals now include uncertainties in the masses of Jupiter,
Uranus, and Neptune and in the orientation of the orbit of Jupiter
(Figure~\ref{fig:lazio.gw}).

Over the next 50~years, improvements in the solar system ephemeris may
be possible, though it is not clear that they will be sufficient to
obtain $\sigma_{\mathrm{SSB}} < 10$~ns.  For instance, connecting data
from the \textit{Galileo} and Juno missions may improve knowledge of
Jupiter's orbit substantially.  Alternately, it may be possible to
incorporate pulsar timing data into determination of the solar system
ephemerides, but such an ephemeris would not be independent for the
purposes of pulsar timing.

\vspace*{-1ex}
\section{Navigation, Pulsar Timing, and the Solar System Ephemeris}\label{sec:lazio.nav}

There has been a long standing interest in (semi-)autonomous
navigation of deep space spacecraft, including the use of X-ray
pulsars (e.g., Chester \& Butman~1981; Sheikh et al.~2006; Deng et
al.~2013; Shemar et al.~2016).  There are even initial tests of the
concept in low-Earth orbit (e.g., Zheng et al.~2017; Neutron star
Interior Composition Explorer/Station Explorer for X-Ray Timing and
Navigation [NICER/SEXTANT]).

In general, considerations of the performance of an X-ray navigation
system have not taken into account uncertainties from knowledge of the
solar system barycenter.
Other considerations also suggest that X-ray navigation is
likely to be of limited use beyond geosynchronous orbit:
\begin{description}
\item[Target body-relative navigation:]~X-ray navigation obtains
positions relative to the barycenter.  While such
positions may be useful during a mission's deep-space cruise,
many missions also require target body-relative navigation.
Examples include the portion of \textit{Cassini}'s Grand Finale
Mission during which it passed only 50~km above the surface of Enceladus
and many small-body missions (e.g., Rosetta at Comet
67P/Churyumov-Gerasimenko, the Hayabusa~2 mission to 
162173~Ryugu, the planned Lucy mission to Jupiter Trojan asteroids,
and the planned Psyche mission to 16~Psyche).  Sheikh et al.~(2006)
speculated on how target body-relative navigation could be
accomplished, but we are unaware of analysis demonstrating that likely
mission requirements could be achieved; Rong et al.~(2016) described a
possible implementation but augmented the X-ray navigation with
an optical camera.

\item[Science measurements:]~Radio navigation has enabled Radio Science.  For instance, a prime
science goal for the Juno mission is to determine the interior
structure of Jupiter (Bolton et al.~2017), which is achieved via
the radio communication-navigation system; a similar study
of Saturn's interior was enabled by \textit{Cassini}'s radio 
communication-navigation system (Edgington \& Spilker~2016).
A typical time scale for orbit determination during Radio Science
measurements is 100~s, and future missions may require measurements
on~10~s time scales.  By contrast, X-ray navigation position
determinations are estimated to require 3000~s or longer (e.g., Shemar
et al.~2016).

\item[Separate communications infrastructure:]~At radio frequencies,
navigation and communications are accomplished with the same
equipment.  While integrated optical communication-navigation payloads
are not yet available, conceptually these functions can be merged
(e.g., Ely \& Seubert~2015), and autonomous optical navigation has
been demonstrated (Deep Space~1, {Rayman} et al.~2000).  Even if an
X-ray communication-navigation payload were developed, it would only
be applicable in free space.  A separate system, at radio or optical
frequencies, would be required for communication to the surface of the
Earth (or through a planetary atmosphere).
\end{description}

\begin{acknowledgements}
Some of the information presented is predecisional information for
planning and discussion only.  
This research has made use of NASA's Astrophysics Data System.
Part of this research was carried out
at the Jet Propulsion Laboratory, California Institute of Technology,
under a contract with the National Aeronautics and Space
Administration.  The NANOGrav project receives support from National
Science Foundation Physics Frontier Center award number 1430284.
\end{acknowledgements}

\end{document}